\documentclass[twoside,journal]{IEEEtran}

\usepackage{amsmath,amssymb,amsthm} 
\usepackage{url}
\usepackage{cite}
\usepackage{graphicx}
\usepackage{hyperref}

\theoremstyle{plain}
\newtheorem{Pro}{Proposition}

\theoremstyle{definition}
\newtheorem{Rem}{Remark}
\newtheorem{Exax}{Example}
\newenvironment{Exa}
  {\pushQED{\qed}\Exax}
  {\popQED\endExax}

\newcommand{\bb}[1]{\mathbb{#1}}
\newcommand{\abs}[1]{\left|{#1}\right|}
\newcommand{\norm}[1]{\left\|{#1}\right\|}
\newcommand{\scalarp}[1]{\left\langle{#1}\right\rangle}
\DeclareMathOperator{\supp}{supp}

\begin{document}
\title{A Phase Vocoder based on Nonstationary \\Gabor Frames}
\author{Emil Solsb{\ae}k Ottosen
		and Monika D{\"o}rfler
		
\thanks{M. D{\"o}rfler has been supported by the Vienna Science and Technology Fund (WWTF) through project MA14-018}		
\thanks{E. S. Ottosen is with the Department of Mathematical Sciences, Aalborg University, 9220 Aalborg, Denmark (e-mail: emilo@math.aau.dk)}
\thanks{M. D{\"o}rfler is with the Faculty of Mathematics, University of Vienna, A-1090 Wien, Austria (e-mail: monika.doerfler@univie.ac.at)}
}

\markboth{IEEE/ACM transactions on audio, speech and language processing}%
{Ottosen \MakeLowercase{\textit{et al.}}: A phase vocoder based on nonstationary Gabor frames}

\maketitle

\begin{abstract}
We propose a new algorithm for time stretching music signals based on the theory of nonstationary Gabor frames (NSGFs). The algorithm extends the techniques of the classical phase vocoder (PV) by incorporating adaptive time-frequency (TF) representations and adaptive phase locking. The adaptive TF representations imply good time resolution for the onsets of attack transients and good frequency resolution for the sinusoidal components. We estimate the phase values only at peak channels and the remaining phases are then locked to the values of the peaks in an adaptive manner. During attack transients we keep the stretch factor equal to one and we propose a new strategy for determining which channels are relevant for reinitializing the corresponding phase values. In contrast to previously published algorithms we use a non-uniform NSGF to obtain a low redundancy of the corresponding TF representation. We show that with just three times as many TF coefficients as signal samples, artifacts such as phasiness and transient smearing can be greatly reduced compared to the classical PV. The proposed algorithm is tested on both synthetic and real world signals and compared with state of the art algorithms in a reproducible manner.
\end{abstract}


\IEEEpeerreviewmaketitle

\section{Introduction}
The task of time stretching or pitch shifting music signals is fundamental in computer music and has many applications within areas such as transcription, mixing, transposition, and auto-tuning \cite{Ishizaki2009,Risset20022}. Time stretching is the operation of changing the length of a signal, without affecting its spectral content, whereas pitch shifting is the operation of raising or lowering the original pitch of a sound without affecting its length. As pitch shifting can be performed by combining time stretching and sampling rate conversion, we shall only focus on time stretching in this paper. 

Introduced by Flanagan and Golden in \cite{Flanagan1966}, the phase vocoder (PV) stretches a signal by modifying its short time Fourier transform (STFT) in such a way that a stretched version can be obtain by reconstructing with respect to a different hop size. Through the years many improvements have been made and the PV is today a well-established technique \cite{Portnoff1976,Griffen1984,Laroche1995,Laroche1999}. Unfortunately, it is known that the PV induces artifacts known as "phasiness" and "transient smearing" \cite{Laroche1999}. Phasiness is perceived as a characteristic colouration of the sound whereas transient smearing is heard as a lack of sharpness at the transients. Many modern techniques exist for dealing with these problems \cite{Laroche1999,Roebel2003,Dorran2004}, but with only few exceptions \cite{Bonada2000,Derrien2007,Liuni2013}, they are all based on the traditional idea of modifying a time-frequency (TF) representation obtained through the STFT. The STFT applies a sampling grid corresponding to a uniform TF resolution over the whole TF plane. For music signals it is often more appropriate to use good time resolution for the onset of attack transients and good frequency resolution for the sinusoidal components. We will consider the task of time stretching in the framework of Gabor theory \cite{Christensen2016,Grochenig2001}. Applying nonstationary Gabor frames (NSGFs) \cite{NuHAGnsgf2011,Dorfler2014} we extend the theory of the PV to incorporate TF representations with the above-mentioned adaptive TF resolution. 

In Section \ref{Sec:stateoftheart} of this article we describe some related work and explain the contributions of the proposed algorithm in relation to state of the art. In Section \ref{Sec:GaborTheory} we introduce the necessary tools from Gabor theory, including the painless condition for NSGFs. We use this framework to present the classical PV in Section \ref{Sec:PhaseVocoder} and the proposed algorithm in Section \ref{Sec:PVNSGF}. We include the derivation of the classical PV for two reasons: Firstly, because it makes the transition to the nonstationary case easier and secondly, because we have not found any other thorough derivation in the literature that uses the framework of Gabor theory. Finally, in Section \ref{Sec:Experiments} we provide the numerical experiments and in Section \ref{Sec:Conclusion} we give the conclusions.

\section{State of the art}\label{Sec:stateoftheart}
Traditionally, time-stretching algorithms are categorized into time-domain and frequency-domain techniques \cite{Laroche1995}. Time-domain techniques, such as \emph{synchronous overlap-add} (SOLA) \cite{Roucos1985} (and its extension PSOLA \cite{Charpentier1986}), are capable of producing good results for monophonic signals, at a low computational cost, but tend to perform poorly when applied to polyphonic signals such as music. 

In contrast, frequency-domain methods, such as the PV \cite{Flanagan1966}, also work for polyphonic signals but with induced artifacts of their own, namely phasiness and transient smearing. As a first improvement to reduce phasiness, Puckette \cite{Puckette1995} suggested to use \emph{phase-locking} to keep phase coherence intact over neighbouring frequency channels. This method was further studied by Laroche and Dolson \cite{Laroche1999} who proposed to separate the frequency axis into \emph{regions of influences}, located around \emph{peak channels}, and to lock the phase values of channels in a given region according to the phase value of the corresponding peak channel. 

To deal with the problem of transient smearing, Bonada \cite{Bonada2000} proposed to keep the stretch factor equal to one during attack transients and then \emph{reinitialize} all phase values for channels above a certain frequency cut, i.e. the phase values of these channels are set equal to the original phase values. In this way, the original timbre is kept intact without ruining the phase coherence for stationary partials at the lower frequencies. A more advanced approach for reducing transient smearing was presented by R{\"o}bel in \cite{Roebel2003}. Here, the transient detection algorithm works on the level of frequency channels and the reinitialization of a detected channel is performed for all time instants influenced by the transient. In this way, there is no need to set the stretch factor equal to one, which is a great advantage in regions with a dense set of transients.

More recent techniques have successfully reduced the PV artifacts by applying more sophisticated TF representations than the STFT. Bonada proposed the application of different FFTs for each time instant, which results in a TF representation with good frequency resolution at the lower frequencies and good time resolution at the higher frequencies. Derrien \cite{Derrien2007} suggested to construct an adaptive TF representation by choosing TF coefficients from a multi-scale Gabor dictionary under a matching constraint. A more recent algorithm, based on the theory of NSGFs, was proposed by Liuni et al. \cite{Liuni2013}. The idea behind their algorithm is to choose a fixed number of frequency bands and to apply, in each band, a NSGF with resolution varying in time. The window functions corresponding to the NSGFs are adapted to the signal by minimizing the \emph{R{\'e}nyi entropy}, which ensures a sparse TF representation. The techniques described in \cite{Roebel2003} and \cite{Liuni2013} are both implemented in the (commercialized) \emph{super phase vocoder} (SuperVP) from IRCAM\footnote{\url{http://anasynth.ircam.fr/home/english/software/supervp}}.

\subsection*{Contributions to state of the art}
In order to generalize the techniques from the classical PV to the case where the TF representation is obtained through a NSGF, it is necessary to use the same number of frequency channels for each time instant. This construction corresponds to a \emph{uniform} NSGF and, since the number of frequency channels must be at least equal to the length of the largest window function, necessarily leads to a high redundancy of the resulting transform. 

In this paper we propose an algorithm, which fully exploits the potential of NSGFs to provide adaptivity while keeping a redundancy similar to the classical PV. This is achieved by letting the number of frequency channels for a given time instant equal the length of the window function selected for that particular time instant. This approach allows for using very long window functions, which is an advantage in regions with stationary partials. We summarize the contributions of this article as follows:
\begin{enumerate}
\item We explain the classical PV and the proposed algorithm in a unified framework using discrete Gabor theory.

\item We present a new time stretching algorithm, which uses an adaptive TF representation of lower redundancy than any other previously published algorithm.

\item While the proposed algorithm combines various familiar techniques from the literature, several new techniques are introduced in order to tackle the challenges arising from the application of non-uniform NSGFs. Hence, the proposed algorithm relies on techniques such as phase locking \cite{Laroche1999}, transient detection \cite{Dixon06}, and quadratic interpolation \cite{Beauregard2015} and integrates new methods for dealing with attack transients (cf. Section \ref{Sec:PVNSGFmod}), for determining the phase values from frequencies estimated by quadratic interpolation (cf. Section \ref{Sec:PVNSGFmod}), and for constructing the stretched signal from the modified (non-uniform) NSGF (cf. Section \ref{Sec:PVNSGFsyn}).

\item We provide a collection of sound files on-line (cf. Section \ref{Sec:Experiments}) and include all source code necessary for reproducing the results.
\end{enumerate}
\section{Discrete Gabor theory}\label{Sec:GaborTheory}
We write $f=(f[0],\ldots,f[L-1])^T$ for a vector $f\in \bb{C}^L$ and $\mathbb{Z}_L=\{0,\ldots,L-1\}$ for the cyclic group. Given $a,b\in \bb{Z}_L$, we define the \emph{translation} operator $\textbf{T}_a:\bb{C}^L\rightarrow \bb{C}^L$ and the \emph{modulation} operator $\textbf{M}_b:\bb{C}^L\rightarrow \bb{C}^L$ by
\begin{equation*}
\textbf{T}_af[l]:=f[l-a]\quad \text{and}\quad \textbf{M}_bf[l]:=f[l]e^{\frac{2\pi i bl}{L}}, 
\end{equation*}
for $l=0,\ldots,L-1$ and with translation performed modulo $L$. For $g\in \bb{C}^L$ and $a,b\in \bb{Z}_L$, we define the \emph{Gabor system} $\{g_{m,n}\}_{m\in \bb{Z}_M,n\in \bb{Z}_N}$ as
\begin{equation*}
g_{m,n}[l]:=\textbf{T}_{na}\textbf{M}_{mb}g[l]=g[l-na]e^{\frac{2\pi imb(l-na)}{L}},
\end{equation*}
with $Na=Mb=L$ for some $N,M\in \bb{N}$ \cite{Strohmer1998,Sondergaard2007}. If $\{g_{m,n}\}_{m,n}$ spans $\bb{C}^L$, then it is called a \emph{Gabor frame}. The associated \emph{frame operator} $\textbf{S}:\bb{C}^L\rightarrow \bb{C}^L$, defined by
\begin{equation*}
\textbf{S}f:=\sum_{m=0}^{M-1}\sum_{n=0}^{N-1}\scalarp{f,g_{m,n}}g_{m,n},\quad \forall f\in \bb{C}^L,
\end{equation*}
is invertible if and only if $\{g_{m,n}\}_{m,n}$ is a Gabor frame\cite{Christensen2016}. If $\textbf{S}$ is invertible, then we have the expansions
\begin{equation}\label{eq:DGTexpansion}
f=\sum_{m=0}^{M-1}\sum_{n=0}^{N-1}\scalarp{f,g_{m,n}}\tilde{g}_{m,n},\quad \forall f\in \bb{C}^L,
\end{equation}
with $\tilde{g}_{m,n}:=\textbf{T}_{na}\textbf{M}_{mb}\textbf{S}^{-1}g$. We say that $\{\tilde{g}_{m,n}\}_{m,n}$ is the \emph{canonical dual frame} of $\{g_{m,n}\}_{m,n}$ and that $\textbf{S}^{-1}g$ is the \emph{canonical dual window} of $g$. The \emph{discrete Gabor transform} (DGT) of $f\in \bb{C}^L$ is the matrix $\textbf{c}\in \bb{C}^{M\times N}$ given by the coefficients $\{\langle f,g_{m,n}\rangle\}_{m,n}$ in the expansion \eqref{eq:DGTexpansion}. Finally, the ratio $MN/L$ is called the \emph{redundancy} of $\{g_{m,n}\}_{m,n}$.
\subsection*{Nonstationary Gabor frames}\label{Sec:NSGFTheory}
\begin{figure}[h!]
\centering
\includegraphics[width=\columnwidth]{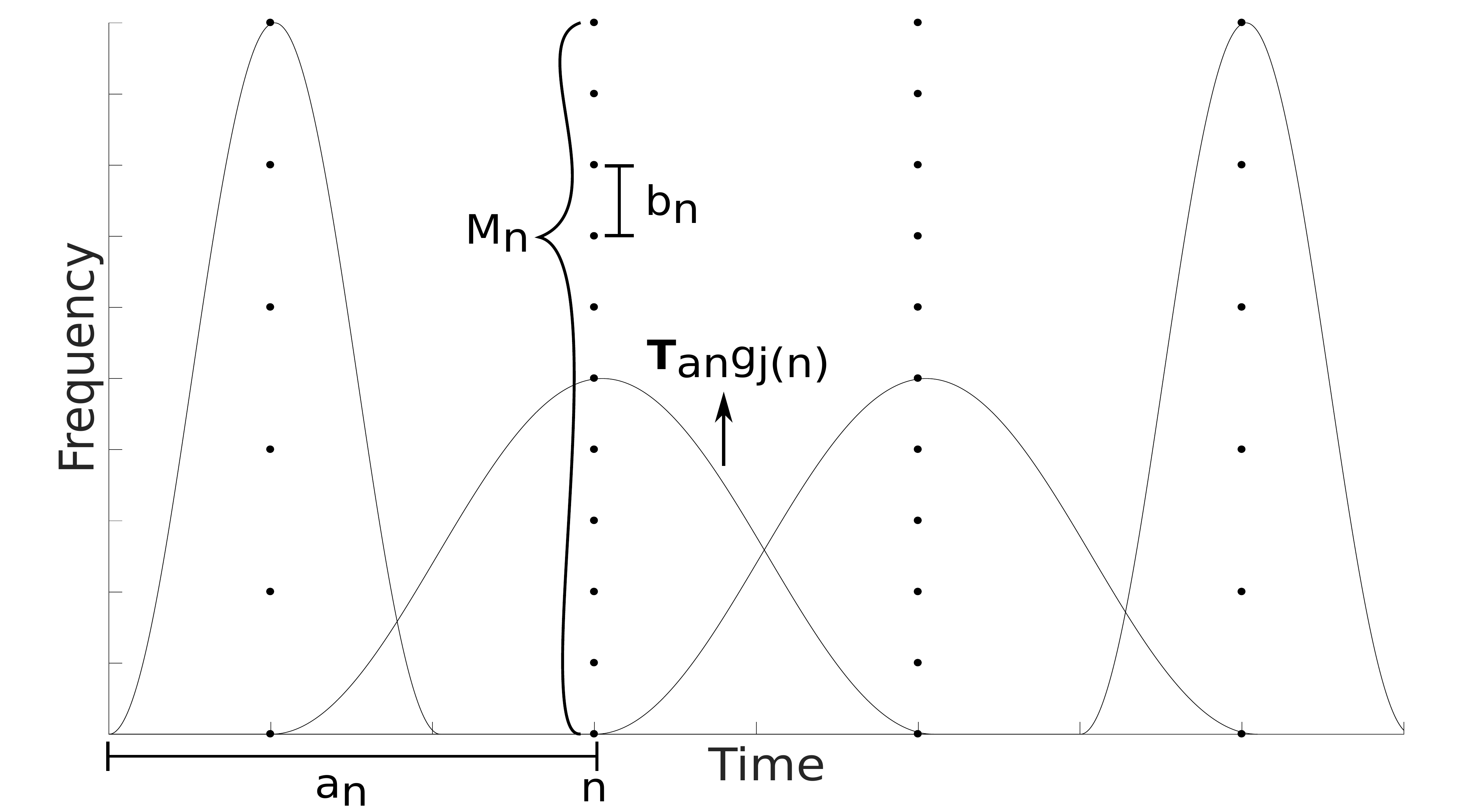}  
\caption{Illustration of a NSGS with $N_w=2$ and $N=4$.}  
\label{Fig:NSGTIllustration}                                                             
\end{figure}

In this section we extend the classical Gabor theory to the nonstationary case \cite{NuHAGnsgf2011}. Just as for the stationary case, we denote the total number of sampling points in time by $N\in \bb{N}$, however, we do not assume these points to be uniformly distributed. Further, instead of using just one window function, we apply $N_w\leq N$ different window functions $\{g_n\}_{n\in \bb{Z}_{N_w}}$ to obtain a flexible resolution. The window function corresponding to time point $n\in \bb{Z}_N$ is denoted by $g_{j(n)}$ with $j:\bb{Z}_N\rightarrow \bb{Z}_{N_w}$ being a surjective mapping. The number of frequency channels corresponding to time point $n\in \bb{Z}_N$ is denoted by $M_n\in \bb{Z}_L$ and the resulting frequency hop size by $b_n:=L/M_n$. Finally, the window functions $\{g_n\}_{n\in \bb{Z}_{N_w}}$ are assumed to be symmetric around zero and we use translation parameters $\{a_n\}_{n\in \bb{Z}_N}\subset \bb{Z}_L$ to obtain to the proper support. With this notation, the \emph{nonstationary Gabor system} (NSGS) $\{g_{m,n}\}_{m\in \bb{Z}_{M_n},n\in \bb{Z}_N}$ is defined as
\begin{equation*}
g_{m,n}[l]:=\textbf{T}_{a_n}\textbf{M}_{mb_n}g_{j(n)}[l]=g_{j(n)}[l-a_n]e^{\frac{2\pi i mb_n (l-a_n)}{L}}.
\end{equation*}
If $\{g_{m,n}\}_{m,n}$ spans $\bb{C}^L$, then it is called a NSGF. If $M_n:=M$, for all $n\in \bb{Z}_N$, then it is called a uniform NSGS (or uniform NSGF if it is also a frame). In Fig. \ref{Fig:NSGTIllustration} we see an example of a simple (non-uniform) NSGS with $N_w=2$ and $N=4$.

Let us now show that the theory of NSGFs extends the theory of standard Gabor frames.
\begin{Exa}
Let $g\in \bb{C}^L$ and $a,b\in \bb{Z}_L$ satisfy $Na=Mb=L$ for some $N,M\in \bb{N}$. Then, with $g_{j(n)}:=g$, $a_n:=na$, and $b_n:=b$ for all $n\in \bb{Z}_N$, we obtain the NSGS
\begin{equation*}
g_{m,n}[l]=\textbf{T}_{na}\textbf{M}_{mb}g[l],\quad m\in \bb{Z}_{M},\quad n\in\bb{Z}_N,
\end{equation*}
which just corresponds to a standard Gabor system.
\end{Exa}
The total number of elements in a NSGS $\{g_{m,n}\}_{m,n}$ is given by $P=\sum_{n=0}^{N-1}M_n$ and the redundancy is therefore $P/L$. The associated frame operator $\textbf{S}:\bb{C}^L\rightarrow \bb{C}^L$, defined by
\begin{equation*}
\textbf{S}f:=\sum_{n=0}^{N-1}\sum_{m=0}^{M_n-1}\scalarp{f,g_{m,n}}g_{m,n},\quad \forall f\in \bb{C}^L,
\end{equation*}
is invertible if and only if $\{g_{m,n}\}_{m,n}$ constitutes a NSGF. If $\textbf{S}$ is invertible, then we have the expansions
\begin{equation}\label{eq:NSGTexpansion}
f=\sum_{n=0}^{N-1}\sum_{m=0}^{M_n-1}\scalarp{f,g_{m,n}}\tilde{g}_{m,n},\quad \forall f\in \bb{C}^L,
\end{equation}
with $\{\tilde{g}_{m,n}\}_{m,n}:=\{\textbf{S}^{-1}g_{m,n}\}_{m,n}$ being the canonical dual frame of $\{g_{m,n}\}_{m,n}$. The \emph{nonstationary Gabor transform} (NSGT) of $f\in \bb{C}^L$ is given by the coefficients $\{c\{n\}(m)\}_{m,n}:=\{\langle f,g_{m,n}\rangle\}_{m,n}$ in the expansion \eqref{eq:NSGTexpansion}. We note that these coefficients do not form a matrix in the general case. We now consider an important case for which the calculation of $\{\tilde{g}_{m,n}\}_{m,n}$ is particularly simple.

\paragraph*{Painless NSGFs} If $\supp(g_{j(n)})\subseteq [c_{j(n)},d_{j(n)}]$ and $d_{j(n)}-c_{j(n)}\leq M_n$ for all $n\in \bb{Z}_n$, then $\{g_{m,n}\}_{m,n}$ is called a \emph{painless} NSGS (or painless NSGF if it is also a frame). In this case we have the following result \cite{NuHAGnsgf2011}.
\begin{Pro}\label{Pro:PainlessNSGT}
If $\{g_{m,n}\}_{m,n}$ is a painless NSGS, then the frame operator $\textbf{S}$ is an $L\times L$ diagonal matrix with entries
\begin{equation*}
S_{ll}=\sum_{n=0}^{N-1}M_n\abs{g_{j(n)}[l-a_n]}^2,\quad \forall l\in \bb{Z}_L.
\end{equation*}
The system $\{g_{m,n}\}_{m,n}$ constitutes a frame for $\bb{C}^L$ if and only if $\sum_{n=0}^{N-1}M_n\abs{g_{j(n)}[l-a_n]}^2>0$ for all $l\in \bb{Z}_L$, and in this case the canonical dual frame $\{\tilde{g}_{m,n}\}_{m,n}$ is given by
\begin{equation*}
\tilde{g}_{m,n}[l]=\frac{g_{m,n}[l]}{\sum_{n'=0}^{N-1}M_{n'}\abs{g_{j(n')}[l-a_{n'}]}^2},
\end{equation*}
for all $n\in \bb{Z}_N$ and all $m\in \bb{Z}_{M_n}$.
\end{Pro}
We note that the canonical dual frame is also a painless NSGF, which is a property not shared by general NSGFs. An immediate consequence of Proposition \ref{Pro:PainlessNSGT} is the classical result for painless nonorthogonal expansions \cite{Daubechies1986}, which just corresponds to the painless case for standard Gabor frames.
\section{The phase vocoder}\label{Sec:PhaseVocoder}
In this section we explain the classical PV \cite{Laroche1999} in the framework of Gabor theory. The PV stretches the length of a signal by means of modifying its discrete STFT. Since the discrete STFT corresponds to a DGT, this technique can be perfectly well explained using Gabor theory. The main idea is to construct a DGT of the signal with respect to an \emph{analysis} hop size $a$, modifying the DGT, and then reconstructing from the modified DGT using a different \emph{synthesis} hop size $a_*$. We only consider the case $a_*=ra$ for a constant modification rate $r>0$. The case $r>1$ corresponds to slowing down the signal by extending its length whereas $r<1$ corresponds to speeding it up by shortening its length. The PV is a classic analysis-modification-synthesis technique, and we will explain each of these three steps separately in the following sections.
\subsection{Analysis}\label{Sec:PhaseVocoderana}
Let $\{g_{m,n}\}_{m,n}$ be a painless Gabor frame for $\bb{C}^L$. Given a real valued signal $f\in \bb{R}^{L}$, we calculate the DGT $\textbf{c}\in \bb{C}^{M\times N}$ of $f$ with respect to $\{g_{m,n}\}_{m,n}$ as
\begin{equation}\label{eq:DGTcoefficients}
c_{m,n}=\scalarp{f,g_{m,n}}=\sum_{l=0}^{L-1}f[l]\overline{g[l-na]}e^{\frac{-2\pi imb(l-na)}{L}},
\end{equation}
for all $m\in \bb{Z}_{M}$ and $n\in \bb{Z}_N$. Let us explain the consequences of the phase convention used in \eqref{eq:DGTcoefficients}. Define $\Omega_m:=2\pi m/M$ as the center frequency of the $m$'th channel and assume that $g$ is real and symmetric around zero. Then, since $\{g_{m,n}\}_{m,n}$ is painless and $b/L=1/M$, we may write \eqref{eq:DGTcoefficients} as
\begin{equation}\label{eq:Gaborconvolution}
c_{m,n}=\sum_{l=0}^{M-1}f[l]g[na-l]e^{-i\Omega_m(l-na)}=e^{i\Omega_m na}\left(f_m\ast g\right)[na],
\end{equation}
with $f_m[l]:=f[l]e^{-i\Omega_ml}$. If $g$ and $\hat{g}$ are both well-localized around zero, the convolution in \eqref{eq:Gaborconvolution} extracts the \emph{baseband} spectrum of $f_m$ at time $na$. Recalling that $f_m$ is just a version of $f$ that has been modulated down by $m$, this baseband spectrum corresponds to the spectrum of $f$ in a neighbourhood of frequency $m$ at time $na$. Finally, modulating back by $m$ we obtain the \emph{bandpass} spectrum of $f$ in a neighbourhood of frequency $m$ at time $na$. This phase convention is the traditional one used in the PV \cite{Laroche1995,Laroche1999,Robel2013}.
\subsection{Modification}\label{Sec:Mod}
To explain the modification step of the PV, we refer to a quasi-stationary sinusoidal model that $f$ is assumed to satisfy \cite{Laroche2002,McAulay1986}. This model is not used explicitly anywhere in the derivation of the PV, but it serves an important role for explaining the underlying ideas. We assume that $f$ can be written as a \emph{finite} sum of sinusoids
\begin{equation}\label{eq:sinusoidmodel}
f(t)=\sum_{k}A_k(t)e^{i\theta_k(t)},
\end{equation}
in which $A_k(t)$ is the \emph{amplitude}, $\theta_k(t)$ is the \emph{phase}, and $\theta'_k(t)$ is the \emph{frequency} of the $k$'th sinusoid at time $t$. Since the model is quasi-stationary, $A_k(t)$ and $\theta'_k(t)$ are assumed to be slowly varying functions. In particular, they are assumed to be almost constant over the duration of $g$. Based on \eqref{eq:sinusoidmodel}, the perfectly stretched signal $f_*$ at time $na_*=nra$ is given by
\begin{equation}\label{eq:timescaledversion}
f_*[na_*]=\sum_{k}A_k(na)e^{ir\theta_k(na)}.
\end{equation}
We note that the amplitudes and frequencies of the stretched signal $f_*$ at time $na_*$ equal the amplitudes and frequencies of the original signal $f$ at time $na$. 

The idea behind the modification step is to construct a new DGT $\textbf{d}\in \bb{C}^{M\times N}$, based on $\textbf{c}\in \bb{C}^{M\times N}$, such that reconstruction from $\textbf{d}$, with respect to $a_*$, yields a time stretched version of $f$ in the sense of \eqref{eq:timescaledversion}. Since the amplitudes need to be preserved we set
\begin{equation*}
d_{m,n}=\abs{c_{m,n}}e^{i\angle d_{m,n}} ,\quad m\in \bb{Z}_M,\quad n\in \bb{Z}_N,
\end{equation*}
using polar coordinates. Estimating the phases $\{\angle d_{m,n}\}_{m,n}$ involves a task called \emph{phase unwrapping} \cite{Laroche1999}.

\paragraph*{Phase unwrapping}
Assume there is a sinusoid of frequency $\omega$ in the vicinity of channel $m$ at time $na$. Then, we make the estimate
\begin{equation}\label{eq:actualphase}
e^{i\angle d_{m,n}}=e^{i(\angle d_{m,n-1}+\omega a_*)},
\end{equation}
since the two DGT samples $d_{m,n-1}$ and $d_{m,n}$ are $a_*$ time samples apart. Using the same argument we may write $e^{i\angle c_{m,n}}=e^{i\left(\angle c_{m,n-1}+\omega a\right)}$. Setting $\omega=\Delta\omega+\Omega_m$, and isolating the deviation $\Delta\omega$, yields
\begin{equation*}
\text{princarg}\left\{\Delta\omega a\right\}=\text{princarg}\left\{\angle c_{m,n}-\angle c_{m,n-1}-\Omega_ma\right\},
\end{equation*}
with "princarg" denoting the principal argument in the interval $]-\pi,\pi]$. Assuming $\omega$ is close to the center frequency $\Omega_m$, such that $\Delta\omega\in ]-\pi/a,\pi/a]$, we arrive at
\begin{equation*}
\Delta\omega=\frac{\text{princarg}\left\{\angle c_{m,n}-\angle c_{m,n-1}-\Omega_m a\right\}}{a}.
\end{equation*}
We can now calculate $\omega$ as $\Delta\omega + \Omega_m$ and use \eqref{eq:actualphase} to determine $\{\angle d_{m,n}\}_{m,n}$ by initializing $d_{m,0}=c_{m,0}$ for all $m\in \bb{Z}_M$.
\subsection{Synthesis}\label{Sec:PhaseVocodersyn}
The final step of the PV is to construct a time stretched version of $f$ in the sense of \eqref{eq:timescaledversion} from the modified DGT $\textbf{d}\in \bb{C}^{M\times N}$. This is done by reconstructing from $\textbf{d}$ with respect to the synthesis hop size $a_*$. According to \eqref{eq:DGTexpansion}, such a reconstruction yields
\begin{equation}\label{eq:synthesisformulagabormodified}
f_*[l]=\sum_{m=0}^{M-1}\sum_{n=0}^{N-1}d_{m,n}\textbf{T}_{na_*}\textbf{M}_{mb}\textbf{S}_*^{-1}g[l],
\end{equation}
with $\textbf{S}_*:\bb{C}^L\rightarrow \bb{C}^L$ being the modified frame operator 
\begin{equation*}
\textbf{S}_*x[l]=\sum_{m=0}^{M-1}\sum_{n=0}^{N_*-1}\scalarp{x,\textbf{T}_{na_*}\textbf{M}_{mb}g}\textbf{T}_{na_*}\textbf{M}_{mb}g[l],	
\end{equation*}
where $N_*:=L/a_*$. The length of the reconstructed signal $f_*$ is given by $L_*=Na_*=Lr$ and translation is performed modulo $L_*$ in \eqref{eq:synthesisformulagabormodified}. In practice, the reconstruction formula \eqref{eq:synthesisformulagabormodified} is realized by applying an inverse FFT and overlap-add.

Traditionally, a DGT with $75\%$ overlap is used in the analysis step, which allows for modification factors $r\leq 4$. We note that if no modifications are made ($r=1$), we recover the original signal. In the next section we consider some of the problems connected with the PV.
\subsection{Drawbacks}\label{Sec:Drawbacks}
The idea behind the PV is intuitive and easily implementable, which makes it attractive from a practical point of view. Unfortunately, the assumptions made in the modification part are not easily satisfied. This is true even for signals constructed explicitly from the sinusoidal model \eqref{eq:sinusoidmodel}. We now list three main problems to be considered.
\begin{enumerate}
\item \textbf{Vertical coherence:} The PV ensures \emph{horizontal coherence} \cite{Laroche1999} within each frequency channel but no attempt is made to ensure \emph{vertical coherence} \cite{Laroche1999} across the frequency channels. If a sinusoid moves from one channel to another, the corresponding phase estimate might change dramatically. This is undesirable since a small change in frequency should only imply a small change in phase.

\item \textbf{Resolution:} In practice, we cannot assume that the sinusoids constituting $f$ are well resolved in the DGT in the sense at most one sinusoid is present within each frequency channel. The channels will only provide a "blurred" image of various neighbouring sinusoids. Furthermore, the amplitudes and frequencies of each sinusoid will often not be constant over the entire duration of $g$. As a consequence, the estimates made in the modification part will be subject to error.

\item \textbf{Transients:} The presence of attack transients is not well modelled within the PV as the phase values at such time instants cannot be predicted from previous estimates. Also, for music signals we often want the attack transients to stay intact after time stretching, which is not accounted for in the PV approach. 
\end{enumerate}
In the next section we construct a new PV, which addresses the above-mentioned problems.
\section[A phase vocoder based on nonstationary Gabor frames]{A phase vocoder based on nonstationary \\Gabor frames}\label{Sec:PVNSGF}
As mentioned in the introduction, the DGT is not always preferable for representing music signals as it corresponds to a uniform resolution over the whole TF plane. A poor TF resolution conflicts with the fundamental idea of well resolved sinusoids and therefore causes problems for the PV. In this section we change the TF representation from the DGT to an adaptable NSGT, which better matches the sinusoidal model \eqref{eq:sinusoidmodel}. To be consistent with the description of the PV in Section \ref{Sec:PhaseVocoder} we separately explain the analysis, modification, and synthesis steps of the proposed algorithm.
 
\subsection{Analysis}\label{Sec:PVNSGFana}
First of all, an adaptation procedure must be chosen for the NSGT. We choose to work with the procedure described in \cite{NuHAGnsgf2011} since it is suitable for representing signals, which consist mainly of transient and sinusoidal components. The adaptation procedure is based on the idea that window functions with small support should be used around the onsets of attack transients whereas window functions with longer support should be used between these onsets. 
\begin{Rem} The construction presented here necessarily yields the problem of a coarse frequency resolution for the transient regions. However, as we propose to keep the stretch factor equal to one during attack transients (cf. Section \ref{Sec:PVNSGFmodtrans}), the impact of this problem is limited.
\end{Rem}
The onsets are calculated using a separate algorithm \cite{Dixon06} and the window functions are constructed as scaled versions of a single window prototype (a Hanning window or similar). The resulting system is referred to as a \emph{scale frame}. In the following paragraphs we explain the construction of scale frames in details.

\paragraph*{Transient detection}
To perform the transient detection we use a spectral flux (SF) onset detection function as described in \cite{NuHAGnsgf2011,Dixon06}. This function is computed with a DGT of redundancy 16, and it measures the sum of (positive) change in magnitude for all frequency channels. A time instant, corresponding to a local maximum of the SF function, is determined as an onset if its SF value is larger than the SF mean value in a certain neighbourhood of time frames. Hence, for region with a dense set of transients, only the most significant onsets are calculated. It is clear that such an approach must be taken to avoid an undesirably low frequency resolution in such regions. The redundant DGT used for the SF onset function is not used anywhere else in our algorithm and does not contribute significantly to the overall complexity.

\paragraph*{Constructing the window functions}\label{Sec:PVNSGFanawindow}
After a set of onsets has been extracted, the window functions are constructed following the rule that the space between two onsets is spanned in such a way that the window length first increases (as we get further away from the first onset) and then decreases (as we approach the next onset). The construction is performed in a smooth way such that the change from one step to the next corresponds to a window function that is either half as long, twice as long or of the same length. For details see \cite{NuHAGnsgf2011}. The overlap between the window functions is chosen such that at most one onset is present within each time frame, we shall elaborate further on this particular construction in Section \ref{Sec:PVNSGFsyn}.

\paragraph*{Constructing the NSGT} Once the window functions $\{g_n\}_{n\in \bb{Z}_{N_w}}$ have been constructed, we choose the numbers of frequency channels $\{M_n\}_{n\in \bb{Z}_N}$ such that the resulting system constitutes a painless NSGF. Additionally, we choose a lower bound on $\{M_n\}_{n\in \bb{Z}_N}$ to avoid an undesirably low number of channels around the onsets (explicit choices of parameters are described in Section \ref{Sec:Experiments}). Given a real valued signal $f\in \bb{R}^L$, we calculate the NSGT $\{c\{n\}(m)\}_{m\in \bb{Z}_{M_n},n\in \bb{Z}_N}$ of $f$ with respect to the scale frame $\{g_{m,n}\}_{m,n}$ as
\begin{equation*}
c\{n\}(m)=\scalarp{f,g_{m,n}}=\sum_{l=0}^{L-1}f[l]\overline{g_{j(n)}[l-a_n]}e^{\frac{-2\pi imb_n(l-a_n)}{L}},
\end{equation*}
for all $n\in \bb{Z}_N$ and all $m\in \bb{Z}_{M_n}$. We note that the phase convention is the same as used in the PV (cf. Section \ref{Sec:PhaseVocoderana}).

\subsection{Modification}\label{Sec:PVNSGFmod}
The idea behind the modification step is the same as for the PV. We assume $f$ satisfies \eqref{eq:sinusoidmodel}, and we construct a modified NSGT $\{d\{n\}(m)\}_{m,n}$, based on $\{c\{n\}(m)\}_{m,n}$, such that reconstruction from $\{d\{n\}(m)\}_{m,n}$, with respect to a set of synthesis translation parameters, yields a time stretched version of $f$ in the sense of \eqref{eq:timescaledversion}. Given a stretch factor $r>0$, the distance between synthesis time sample $n$ and $n+1$ is
\begin{equation}\label{eq:synthhopsize}
a^*_n:=r(a_{n+1}-a_{n}),\quad n\in \bb{Z}_N.
\end{equation}
Since we do not want the transients to be stretched, we let $r=1$, when $a_n$ corresponds to the onset of a transient, and then stretch with a correspondingly larger factor $r'>r$ in remaining regions. Using polar coordinates we set
\begin{equation*}
d\{n\}(m)=\abs{c\{n\}(m)}e^{i\angle d\{n\}(m)} ,\quad n\in \bb{Z}_N,\quad m\in \bb{Z}_{M_n},
\end{equation*}
with $\angle d\{0\}(m)=\angle c\{0\}(m)$ for all $m\in \bb{Z}_{M_0}$. Hence, in complete analogy with the approach in the PV, the problem boils down to estimating the phase values $\{\angle d\{n\}(m)\}_{m,n}$. 

\begin{figure}[h!]
\centering
\includegraphics[width=\columnwidth]{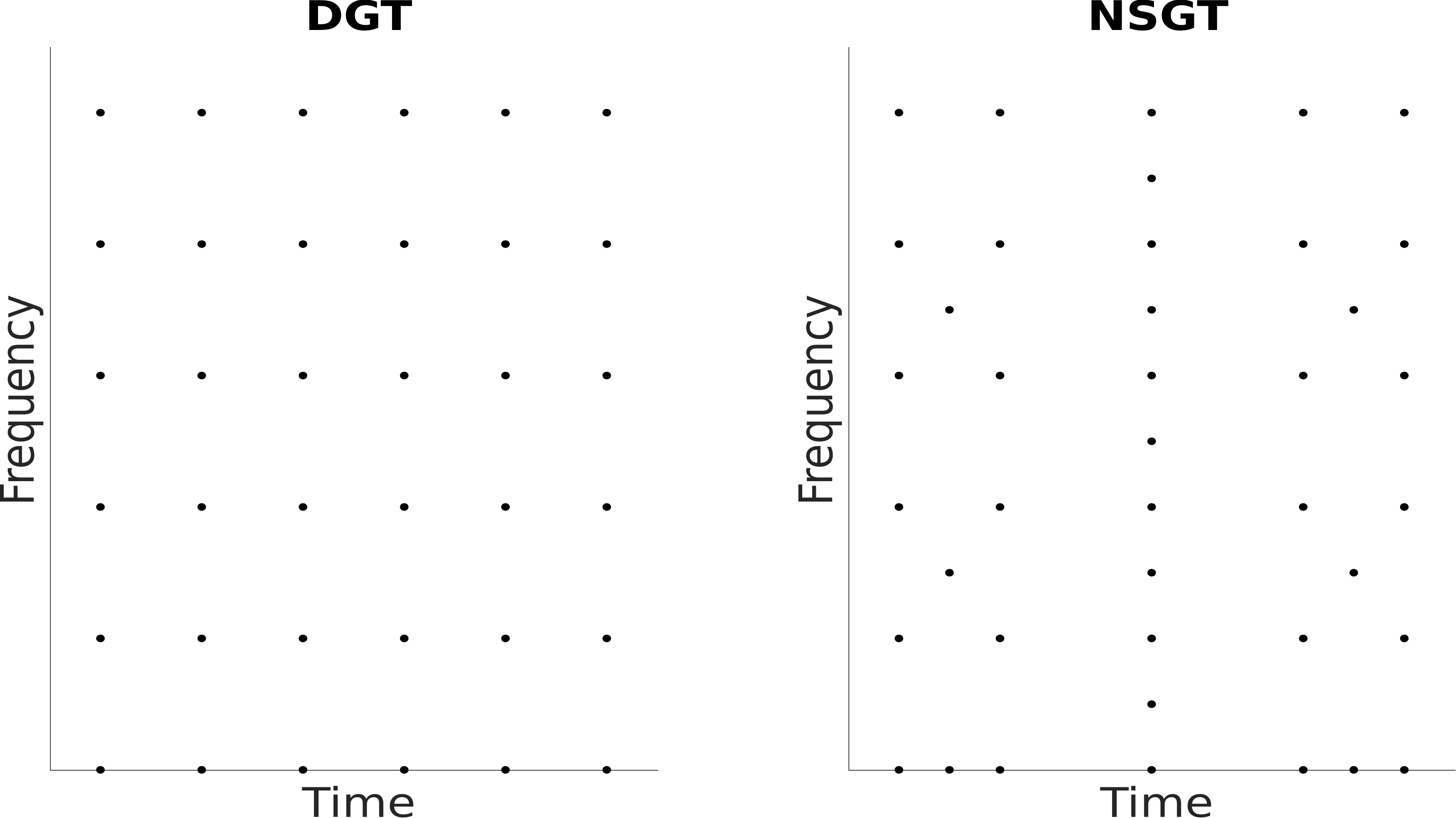}    
\caption{Sampling grids corresponding to a DGT and a NSGT.}  
\label{Fig:Grid}                                                              
\end{figure}

Making the transition from stationary Gabor frames to NSGFs, we are facing a fundamental problem. The DGT corresponds to a uniform sampling grid over the TF plane, whereas the NSGF corresponds to a sampling grid which is irregular over time but regular over frequency for each fixed time position. This is illustrated in Fig. \ref{Fig:Grid}.

As a consequence, we cannot guarantee that each sampling point has a horizontal neighbour that can be used for estimating the frequency as in the PV (cf. Section \ref{Sec:PhaseVocoder}). We therefore generalize the approach from \cite{Beauregard2015} to the nonstationary case and calculate the frequencies using \emph{quadratic interpolation}. 

\begin{figure}[h!]
\centering
\includegraphics[width=\columnwidth]{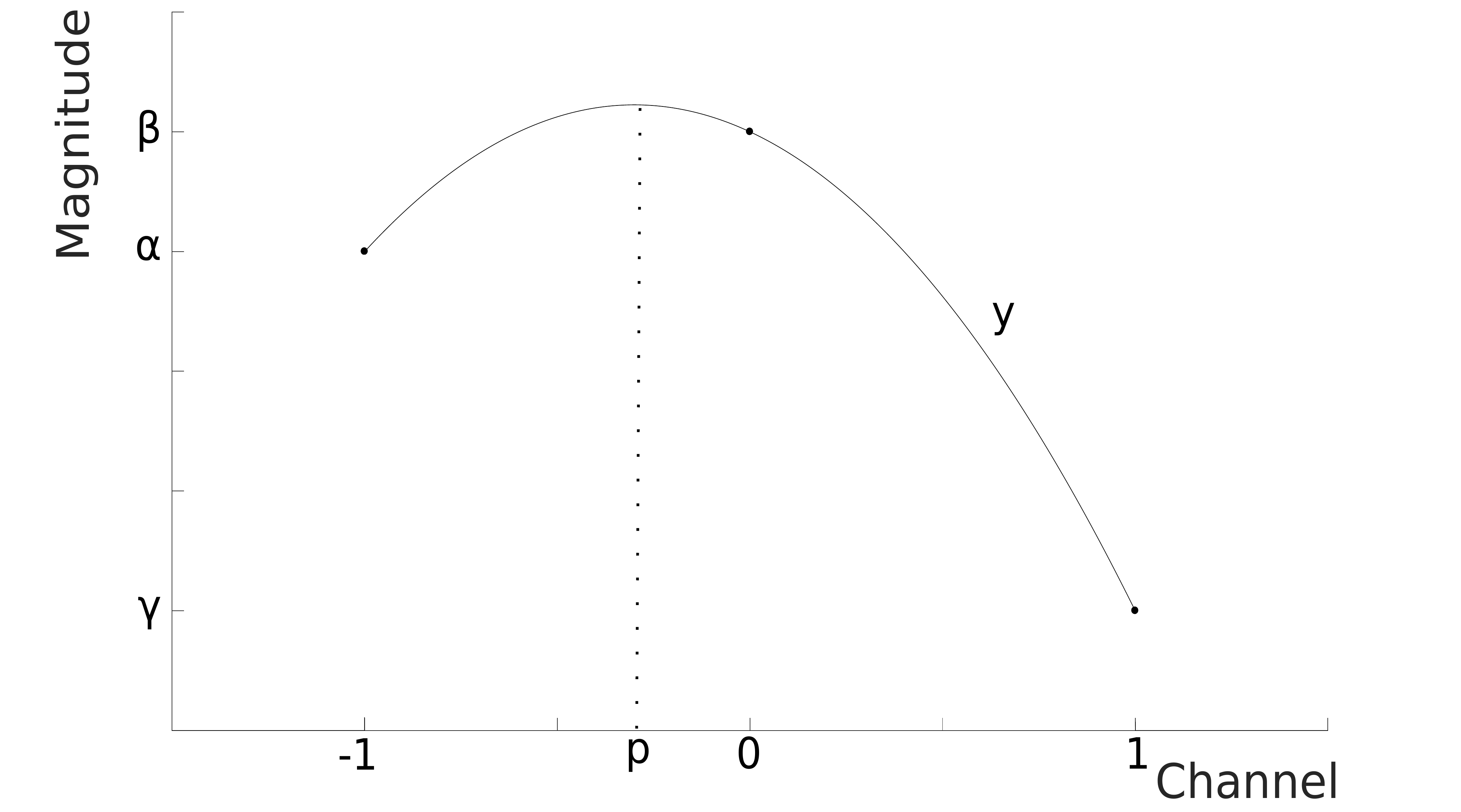}    
\caption{Illustration of quadratic interpolation.}  
\label{Fig:QIFigure}                                                              
\end{figure}

\paragraph*{Calculating the frequencies}
For fixed $n\in \bb{Z}_N$, we define channel $m_p$ as a \emph{peak} if its magnitude $\abs{c\{n\}(m_p)}$ is larger than the magnitudes of its two vertical neighbours, i.e. $\abs{c\{n\}(m_p)}>\abs{c\{n\}(m_p\pm 1)}$. If there is a sinusoid of frequency $\omega$ in the vicinity of peak channel $m_p$, the "true" peak position will differ from $m_p$ unless $\omega$ is exactly equal to $2\pi m_p/M_n$. The idea is thus to interpolate the true peak position, using the neighbouring channels $m_p\pm 1$, and then to apply this value as an estimate for $\omega$. To describe the setup we set the position of the peak channel $m_p$ to $0$, and the positions of its two neighbours to $-1$ and $1$, respectively. Also, we denote the true peak position by $p$ and define
\begin{equation*}
\alpha:=\abs{c\{n\}(-1)}, \quad \beta:=\abs{c\{n\}(0)},\quad \text{and} \quad \gamma:=\abs{c\{n\}(1)}.
\end{equation*}
The situation is illustrated in Fig. \ref{Fig:QIFigure}, with $y$ denoting the parabola to be interpolated.

Writing $y(x)=a(x-p)^2+b$ and solving for $p$ yields
\begin{equation*}
p=\frac{1}{2}\cdot \frac{\alpha-\gamma}{\alpha-2\beta+\gamma}\in \left(-\frac{1}{2},\frac{1}{2}\right).
\end{equation*}
The value of $p$ determines the deviation from the peak channel to the true peak proportional to the size of the channel. After $p$ has been determined, we calculate the frequency as
\begin{equation}\label{eq:freqestimate}
\omega=\frac{2\pi(m_p+p)}{M_n}.
\end{equation}
In practice, the calculations are done on a dB scale for higher accuracy. Let us now explain how the frequency estimate \eqref{eq:freqestimate} is used to calculate the corresponding phase value $\angle d\{n\}(m_p)$.

\begin{figure}[h!]
\centering
\includegraphics[width=\columnwidth]{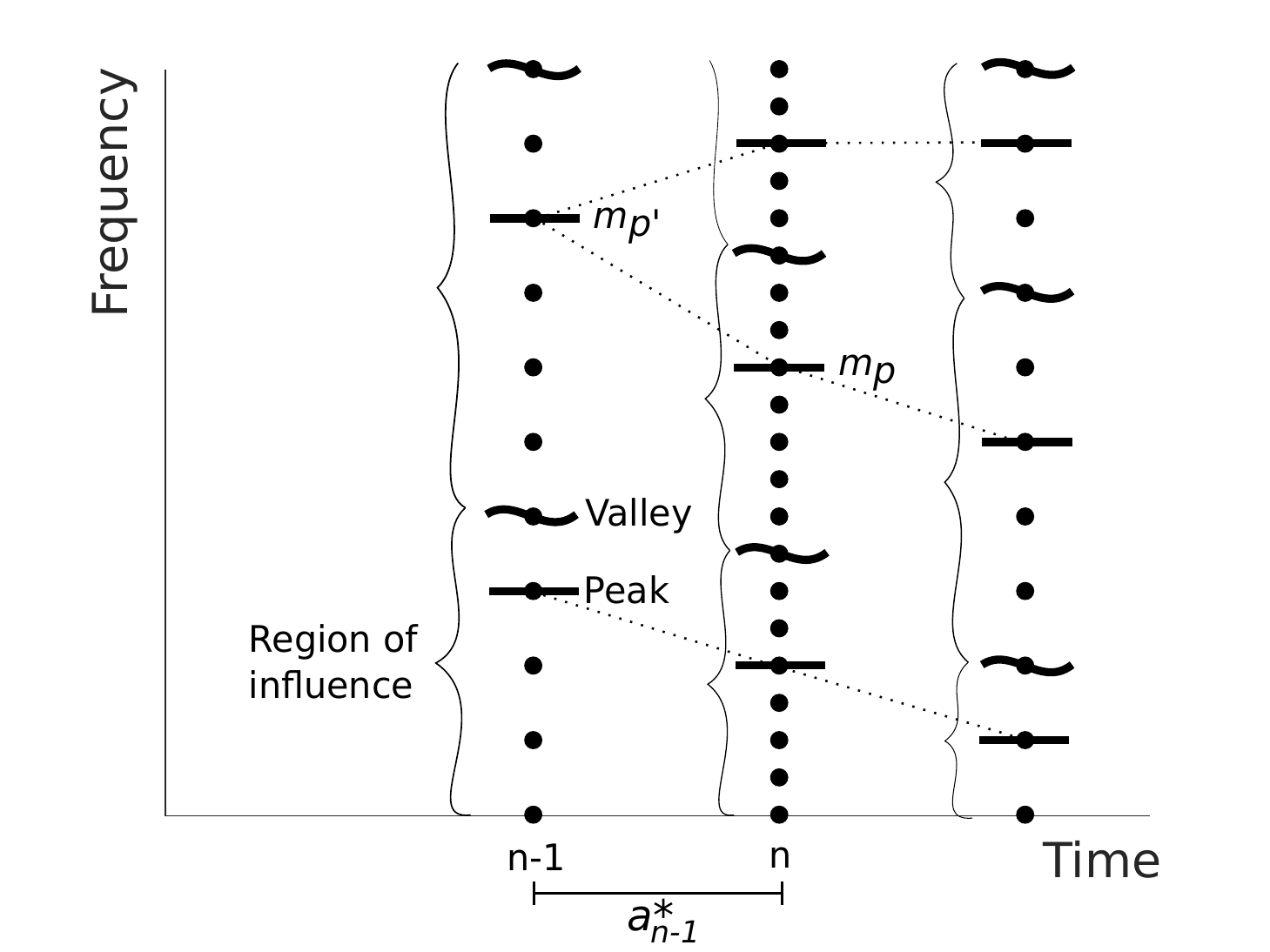}    
\caption{Illustration of peak, valley and region of influence.}  
\label{Fig:RegionOfInfluence}                                                            
\end{figure}

\paragraph*{Calculating the phases}
Between each pair of peaks we define the (lowest) channel with smallest magnitude as a \emph{valley} and then use these valleys to separate the frequency axis into \emph{regions of influence}. As noted in \cite{Laroche1999}, if a peak switches from channel $m_{p'}$ at time $n-1$ to channel $m_p$ at time $n$, the corresponding phase estimate should take this behaviour into account. A simple way of determining the previous peak $m_{p'}$ is to choose the peak of the corresponding region of influence that channel $m_p$ would have belonged to in time frame $n-1$. This is illustrated in Fig. \ref{Fig:RegionOfInfluence}.

Based on this construction, with $a^*_{n-1}$ given in \eqref{eq:synthhopsize}, the phase estimate at peak channel $m_p$ is
\begin{equation}\label{eq:peakchannelphase}
d\{n\}(m_p)=\abs{c\{n\}(m_p)}e^{i(\angle d\{n-1\}(m_{p'})+\omega a^*_{n-1})}.
\end{equation}
For the neighbouring channels in the corresponding region of influence, the phase values will be locked to the phase of the peak. Following the approach in \cite{Laroche1999}, we let
\begin{equation*}
e^{i\angle d\{n\}(m)}=e^{i\left(\angle d\{n\}(m_p)+\angle c\{n\}(m)-\angle c\{n\}(m_p)\right)},
\end{equation*}
for all channels $m$ in the region of influence corresponding to peak channel $m_p$. Hence, the phase locking is such that the difference in synthesis phase is the same as the difference in analysis phase. It is important to note that the actual phase estimates are done only at peak channels, which allows for a fast implementation. As mentioned in Section \ref{Sec:Drawbacks}, the phase estimate \eqref{eq:peakchannelphase} is not well suited for modelling attack transients. In the next section we explain our approach for dealing with this problem.

\paragraph*{Transient preservation}\label{Sec:PVNSGFmodtrans}
Since the phase values $\angle d\{n\}(m)$ at transients locations cannot be predicted from previous estimates, one might choose to simply reinitialize all phase values at such locations $\angle d\{n\}(m)=\angle c\{n\}(m)$. However, for stationary partials passing through the transient, such a reinitialization completely destroys the horizontal phase coherence, thereby producing undesirable artifacts in the resulting sound. To deal with this problem, we propose the following rule for phase estimation at transient locations: Assume time-instant $n$ corresponds to the onset of an attack transient. Consider channel $m$, belonging to the region of influence dominated by a peak channel $m_p$, and let $m_{p'}$ denote the peak channel of the region of influence that channel $m_p$ would have belonged to in time frame $n-1$ (same notation as in \eqref{eq:peakchannelphase}, see also Fig. \ref{Fig:RegionOfInfluence}). Then, given a tolerance $\varepsilon>0$, we reinitialize $\angle d\{n\}(m)=\angle c\{n\}(m)$ if and only if 
\begin{equation}\label{eq:transpre}
\abs{c\{n\}(m)}>\abs{c\{n-1\}(m_{p'})}+\varepsilon.
\end{equation}
For the implementation, the calculations are done on a dB scale with $\varepsilon = 2$dB. We note that in contrast to previously proposed techniques for onset reinitialization\cite{Bonada2000,Derrien2007,Roebel2003}, our algorithm has the advantage that it tracks sinusoids \emph{across} frequency channels. 

\subsection{Synthesis}\label{Sec:PVNSGFsyn}
Before we can provide the actual synthesis formula, we need to return to the problem of choosing the overlap between window functions (cf. Section \ref{Sec:PVNSGFanawindow}). Originally, scale frame were invented with the intention of construction adaptive TF representations with a very low redundancy. To ensure a low redundancy, and a stable reconstruction, the overlap between adjacent window functions is chosen as $1/3$ of the length for equal windows and $2/3$ of the length of the shorter window for different windows \cite{NuHAGnsgf2011}.

\begin{figure}[h!]
\centering
\includegraphics[width=\columnwidth]{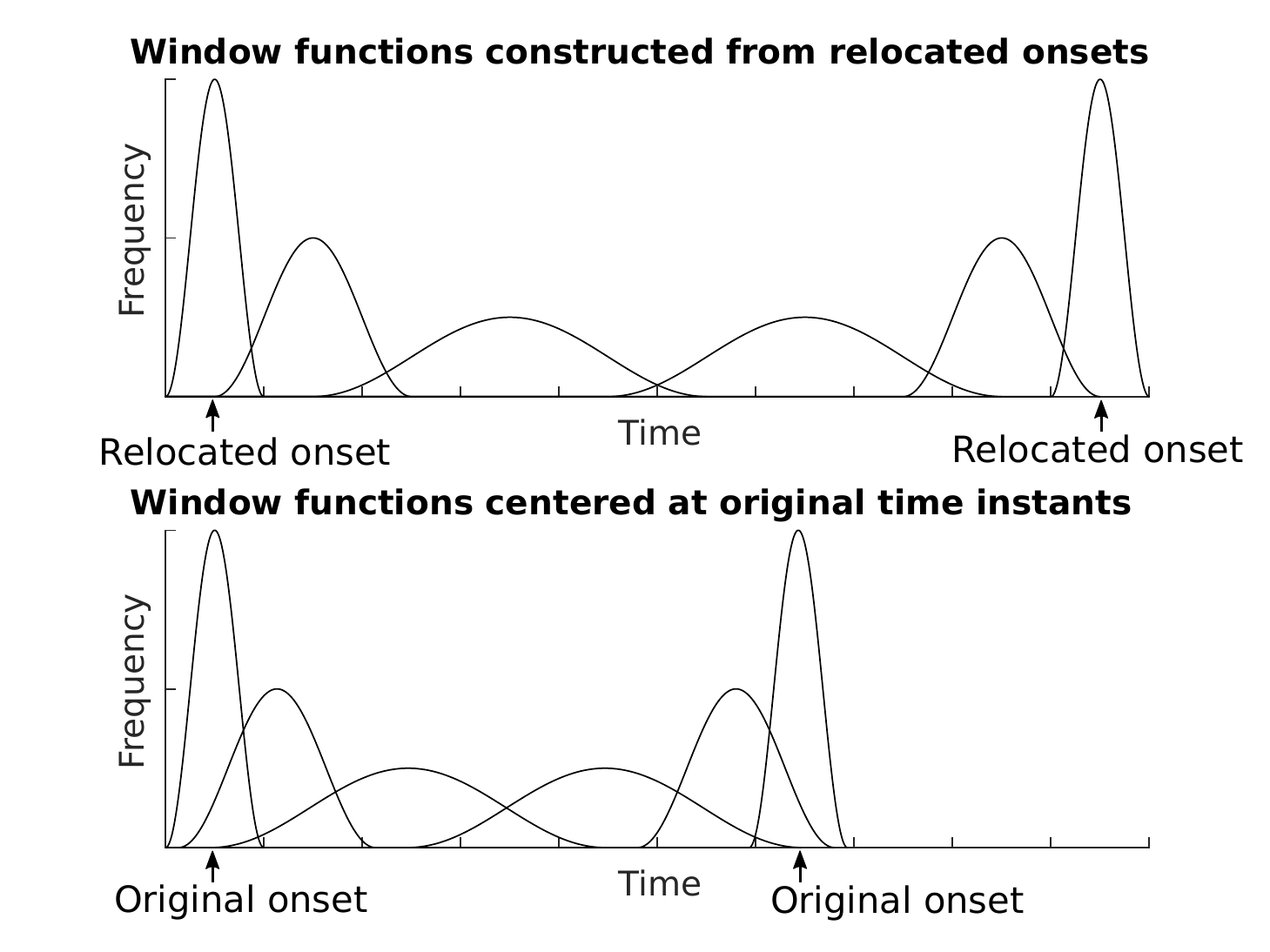}    
\caption{Construction of the overlap between window functions.}  
\label{Fig:WindowFunctions}                                                               
\end{figure}

This construction makes sense in the general settings, since the resulting system constitutes a frame for $\bb{C}^L$ as long as the painless condition from Proposition \ref{Pro:PainlessNSGT} is satisfied. However, in the case of time-stretching with a factor $r>1$, this construction cannot guarantee that the dual windows (cf. Proposition \ref{Pro:PainlessNSGT}) overlap coherently when placed at the synthesis time instants. To tackle this problem, we have chosen the overlap between window functions in the following way:
\begin{enumerate}
\item First the onsets of attack transients are calculated (using the onset detection algorithm from Section \ref{Sec:PVNSGFana}).

\item Then these onsets are relocated such that the distance between the relocated onsets is $r$ times the distance between the original onsets.

\item The window functions are now calculated according to the relocated onsets, using the approach in \cite{NuHAGnsgf2011}, and afterwards centred at the original time instants.
\end{enumerate}
While this approach might give the impression that we just stretch the window lengths by a factor of $r$, this is not the case. Calculating the windows with respect to the relocated onsets still produce a sequence of windows functions of the same lengths as if the original onsets had been used. This is illustrated in Fig. \ref{Fig:WindowFunctions}.

With this choice of overlap, we can construct the stretched signal $f_*$ using the synthesis formula
\begin{equation}\label{eq:synthesisformula}
f_*=\sum_{n\in \bb{Z}_N}\sum_{m\in \bb{Z}_{M_n}}d\{n\}(m)\tilde{g}_{m,n},
\end{equation}
with $\{\tilde{g}_{m,n}\}_{m,n}$ being the canonical dual frame from Proposition \ref{Pro:PainlessNSGT} constructed using the synthesis time instants. In practice, the reconstruction formula \eqref{eq:synthesisformula} is realised by applying an inverse FFT and overlap-add as in the classical PV.

\subsection{Advances}
In this section we explain how the proposed algorithm improves the techniques of the PV. We do so by separately addressing the three drawbacks described in Section \ref{Sec:Drawbacks}.
\begin{enumerate}
\item \textbf{Vertical coherence:} If a sinusoid moves from channel $m_{p'}$ at time $n-1$ to channel $m_p$ at time $n$, then the corresponding peak channel also changes from $m_{p'}$ to $m_p$. The estimate given in \eqref{eq:peakchannelphase} therefore ensures that the corresponding phase increment takes this behaviour into account. In this way we get coherence \emph{across} the various frequency channels in contrast to the standard PV which only provides coherence \emph{within} each frequency channel.

\item \textbf{Resolution:} Changing the representation from that of a DGT to an adaptable NSGT automatically improves the TF resolution for signals, which are well represented by the sinusoidal model \eqref{eq:sinusoidmodel}. Furthermore, calculating the phase increment only at peak channels replaces the underlying assumption of well resolved sinusoids in each frequency channel with the weaker assumption of well resolved sinusoids in each region of influence.

\item \textbf{Transients:} To reduce transient smearing, we keep the stretch factor equal to one during attack transients and we reinitialize the phase values of relevant channels according to \eqref{eq:transpre}.
\end{enumerate}
While the PV serves as a good starting point for understanding the ideas behind the proposed algorithm, it is not the main goal of this article only to improve the resulting sound quality compared to this classical technique. The main advantage of the proposed algorithm is the ability to produce good results, when compared to state of the art, while keeping a low redundancy of the applied TF transform. 

\paragraph*{Redundancy of the NSGT}
As mentioned in Section \ref{Sec:PhaseVocodersyn}, the classical PV applies an overlap of 75$\%$ corresponding to a redundancy of $4$ in the DGT. There is some mathematical justification to this choice \cite{Laroche1999}, but mainly the overlap is chosen to ensure a good TF resolution. It should be noted that the redundancy of the DGT is independent of the signal under consideration --- it only depends on the analysis hop size and the length of the window function (assuming the painless condition is satisfied).

For multi-resolution methods, the situation changes as the TF resolution adapts to the particular signal. A standard approach for multi-resolution methods is to choose non-uniform sampling points in time, with corresponding window functions, and a \emph{uniform} number of frequency channels corresponding to the length of the largest window function \cite{Liuni2013,Laroche1995}. This construction corresponds to applying a uniform NSGF (cf. Section \ref{Sec:NSGFTheory}). Such an approach is desirable from a practical point of view as the coefficients then form a matrix and the standard techniques from the PV (and its improvements) immediately apply. However, the choice of a uniform NSGF naturally implies a high redundancy of the transform as the sampling density is much higher than needed for the painless case (cf. Proposition \ref{Pro:PainlessNSGT}). For real world signals, such a high redundancy is undesirable as it implies a high computational cost for the time-stretching algorithm.

In contrast to previously suggested methods, our algorithm takes full advantage of the painless condition and produces good results with a redundancy of $\approx 3$ for a stretch factor of $r=1.5$. It is important to note that the redundancy of the proposed algorithm depends \emph{both} on the signal under consideration and the stretch factor (at least in the case where $r>1$). For different signals, the onset detection algorithm calculates different onsets, which results in different time sampling points and different numbers of frequency channels. As for the stretch factor, we recall the choice of overlap as described in Section \ref{Sec:PVNSGFsyn}. For a large stretch factor, we need a large overlap between the window functions to guarantee that the synthesis formula \eqref{eq:synthesisformula} makes sense. We do not consider the dependency between the redundancy and the stretch factor a problem, since the redundancy is still manageable even for large stretch factors. For a stretch factor of $r=3$, the redundancy is $\approx 5$ and for a stretch factor of $r=4$, the redundancy is $\approx 7$.

In the next section we present the numerical experiments and compare the proposed algorithm with state of the art algorithms for time stretching (cf. Section \ref{Sec:stateoftheart}).
\section{Experiments}\label{Sec:Experiments}
The proposed algorithm has been implemented in MATLAB R2017A and the corresponding source code is available at
\begin{center}
\url{http://homepage.univie.ac.at/monika.doerfler/NSPV.html}
\end{center}
The source code depends on the following two toolboxes: The LTFAT \cite{ltfatnote030} (version 2.1.2 or above) freely available from \url{http://ltfat.github.io/} and the NSGToolbox \cite{NuHAGnsgf2011} (version 0.1.0 or above) freely available from \url{http://nsg.sourceforge.net/}. 

For the classical PV, we use an implementation by Ellis \cite{Ellis2002}, which includes some improvements to the procedure described in Section \ref{Sec:PhaseVocoder} (in particular, interpolation of magnitudes). As these improvements result in a significantly improved audio quality, we have chosen this implementation for comparison. 

In Section \ref{Sec:Experimentssynth} we compare the proposed algorithm to the classical PV by stretching synthetic (music) signals and in Section \ref{Sec:Experimentsreal} we turn to the analysis of real world signals and compare the proposed algorithm with the algorithms from Derrien \cite{Derrien2007} and Liuni et al. \cite{Liuni2013}.

\subsection{Synthetic signals}\label{Sec:Experimentssynth}
Analysing synthetic signals has the advantage that the perfect stretched version is available and can be used as ground truth. For this experiment, we construct a large number of synthetic signals and compare the performance of the proposed algorithm with the classical PV for each of these signals. More precisely, the approach is as follows:
\begin{enumerate}
\item For each synthetic melody we choose a random number of notes between $4$ and $10$. Each note has a randomly chosen duration of either $0.5$ or $1$ second and the corresponding tone consists of a fundamental frequency and three harmonics of decreasing amplitudes. The fundamental frequencies are set to coincide with those of a piano and the melody is allowed to move either $1$ or $2$ half notes up or down (randomly chosen) per step. A randomly chosen envelope ensures that the tones have both an attack and a release. The sampling frequency of the resulting signal $s$ is $16000$ Hz.

\item A stretch factor $0.5\leq r\leq 3.75$ is chosen at random and another synthetic signal $s_{perf}$ is constructed, such that $s_{perf}$ corresponds to a perfectly time stretched version of $s$ in the sense of \eqref{eq:timescaledversion}. The classical PV and the proposed algorithm are applied to the original signal $s$, with respect to the stretch factor $r$, resulting in the time stretched versions $s_{pv}$ and $s_{nsgt}$.

\item Three DGTs $S_{perf}$, $S_{pv}$, and $S_{nsgt}$ are constructed from the time stretched versions $s_{perf}$, $s_{pv}$, and $s_{nsgt}$ using the same parameter settings for each signal. With $|S|$ denoting a vector consisting of the absolute values of a DGT $S$, we use the following error measure
\begin{equation}\label{eq:SpecMagComp}
E(S_{perf},S)=\frac{\norm{\abs{S_{perf}}-\abs{S}}_2}{\norm{\abs{S_{perf}}}_2},
\end{equation} 
with $S$ being either $S_{pv}$ or $S_{nsgt}$.
\end{enumerate}
Note that we cannot apply a sample by sample error measure in the time domain, since in this case a small change in phase for the stretched signals might cause a large error, which does not reflect the actual sound quality. We therefore choose to compare the stretched versions using the magnitude difference of their DGTs. Let us now define the parameters used for the TF representations in this experiment.
\paragraph*{Choice of parameters}
For the DGT used in the PV, we apply two different parameter settings. Using the notation $(\text{hopsize},\text{number of frequency channels})$ we use the parameters $(256,1024)$ and $(128,512)$. For the first parameter setting we use a Hanning window of length $1024$ and for the second parameter setting we use a Hanning window of length $512$. In this way we obtain painless DGTs of redundancy $4$. 

For the NSGT used in the proposed algorithm, we use $5$ different Hanning windows with lengths varying from $96$ samples (at attack transients) to $96\cdot 2^4=1536$ samples. The lower bound on the number of frequency channels is set to $96\cdot 2^3=768$, corresponding to the length of the second largest window functions. 

For the DGT used for computing $S_{perf}$, $S_{pv}$, and $S_{nsgt}$, we use parameters $(128,2048)$ and a Hanning window of $2048$ samples. This results in a painless DGT of redundancy 16.
\paragraph*{Results}
Repeating the experiment described above for $1000$ synthetic test signals we get the following average results for the redundancy and error of each algorithm:
\begin{table}[h!]
\caption{Average results for $1000$ synthetic test signals}
\label{tab:SpecMagTable}
\centering
\begin{tabular}{l c c c }
\hline
Algorithm: & PV$(256,1024)$ & PV$(128,512)$& Proposed\\
Average red.: & $3.954813$ & $3.977300$ & $3.637370$\\
Average error: & $0.439982$ & $0.415139$ & $0.095104$\\
\hline
\end{tabular}
\end{table}

\begin{figure}[h!]
\centering
\includegraphics[width=\columnwidth]{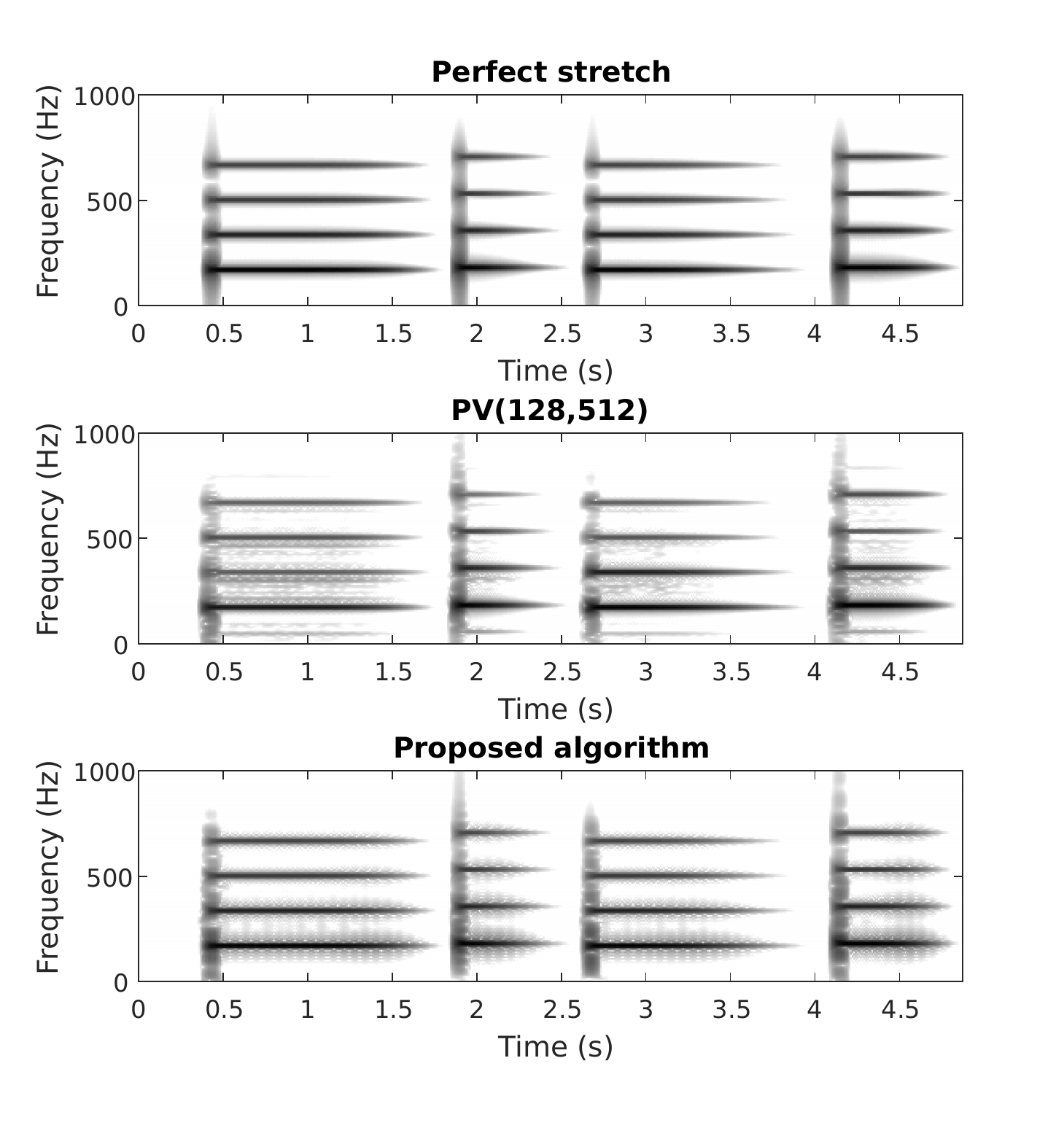}    
\caption{Spectrograms for stretched versions of a synthetic signal with $r=1.5$.}  
\label{Fig:Spectrograms}                                                                
\end{figure}

We note that the proposed algorithm outperforms the classical PV, with respect to the error measure in \eqref{eq:SpecMagComp}, while keeping a comparable redundancy of the applied transform. For a visualization of the performances of the algorithms we have plotted, in Fig. \ref{Fig:Spectrograms}, the spectrograms corresponding to the three DGTs $S_{perf}$, $S_{pv}$ (with parameters $(128,512)$), and $S_{nsgt}$ for one particular synthetic test signal (with $r=1.5$).

We can easily see how the proposed algorithm more accurately reproduces the onsets, and how it reduces the noisy components between the harmonics compared to the PV. However, we can also see how the frequencies corresponding to the harmonics are better reproduced with the PV than with the proposed algorithm. The proposed algorithm induces a certain amplitude modulation due to the peak detection and phase locking approach described in Section \ref{Sec:PVNSGFmod}. 

We have provided sound files on-line for the particular test signal shown in Fig. \ref{Fig:Spectrograms} with respect to the stretch factors $r= [0.75, 1.25, 1.5, 2.25, 3.0, 3.75]$. The sound files are available for the perfect stretched version, the PV$(128,512)$, and the proposed algorithm. It is important to note that the error measure given in \eqref{eq:SpecMagComp} is not a direct reflection of the actual audio quality --- it is for instance not true that the proposed algorithm consistently performs $4$ times as good as the classical PV. The results for the proposed algorithm are particularly convincing for stretch factors $r\leq 2$, where the timbre at attack transients is nicely preserved in contrast to the classical PV. However for larger stretch factors $r\geq 2$, the impact of the amplitude modulation, and of the coarse frequency resolution around onsets, becomes audible. Eventually, this results in an overall sound quality comparable to the PV (or even below for very large stretch factors $r\geq 3$).

Since the authors do not have access to the source code of the more sophisticated algorithms as proposed in \cite{Bonada2000,Derrien2007,Liuni2013}, the comparison for synthetic signals could only be done for the PV and the proposed algorithm. However, as the authors from \cite{Derrien2007} and \cite{Liuni2013} kindly provided us with sound files for real world signals, we have included these algorithms for the comparison in the next section.

\subsection{Real world signals}\label{Sec:Experimentsreal}
For this experiment we consider three real world signals, each of length $\approx 4$ seconds and with a sampling frequency of 44100 Hz. The signals are chosen such that they challenge different aspects of the time stretching algorithms:
\begin{enumerate}
\item The first signal is a glockenspiel signal with few transients and many harmonics at the higher frequencies.

\item The second signal is a piece of piano music consisting of a dense set of transients with most of the energy concentrated at the lower frequencies.

\item The third signal is from a rock song played by a full band, thereby producing a complex polyphonic sound.
\end{enumerate}
We chose to work with the stretch factors 0.75, 1.25, 1.5 and 2.25 for the comparison. The algorithms we include are:
\begin{enumerate}
\item The PV as described in Section \ref{Sec:PhaseVocoder} and implemented in \cite{Ellis2002}. For the DGT used in the PV, we use parameters $(512,2048)$ and a Hanning window of length $2048$.

\item The proposed algorithm from Section \ref{Sec:PVNSGF}. We use $5$ Hanning windows with lengths varying from $384$ to $384\cdot 2^4=6144$ and with $384\cdot 2^2=1536$ being the lower bound on the number of frequency channels.

\item The matching pursuit algorithm by Derrien \cite{Derrien2007}. 

\item The SuperVP from IRCAM based on the theory of R{\"o}bel \cite{Roebel2003} and Liuni et al. \cite{Liuni2013}. The algorithm uses only one frequency band and chooses between window lengths of 1024, 2048, 3072, and 4096 samples for the adaptive (uniform) NSGT. We refer the reader to \cite{Liuni2013} for details.
\end{enumerate}
Since all the stretched sounds are available on-line, we only give the main conclusions. The classical PV and the algorithm by Derrien are rather similar in performance --- they both produce a good overall sound quality but with significant transient smearing. The proposed algorithm, on the other hand, does a much better job of preserving the original timbre at attack transient, but induces a certain "roughness" to the sounds (mainly for $r=2.25$). Also, some of the weaker transients, which are not detected by the onset detection algorithm, suffer from transient smearing for the proposed algorithm (in particular, the "tapping" noises in the background of the piano music). The SuperVP does not have this problem as the transient detection algorithm works on the level of spectral bins. Overall, the SuperVP provides the best audio quality for the three signals, which is to be expected as it applies a TF representation of much higher redundancy than the other algorithms. Calculating the average redundancies for the proposed algorithm (over the $4$ stretch factors) for each signal we get $2.40$, $2.90$ and $2.65$. Finally, let us note that the third signal (the rock band signal) reveals a fundamental problem with the application of NSGFs. For $r=2.25$, neither the proposed algorithm nor the SuperVP are capable of maintaining a steady bass, which results from the changing window lengths. This particular issue is better resolved by the classical PV as well as the algorithm by Derrien.

\section{Conclusion and perspectives}\label{Sec:Conclusion}
Using discrete Gabor theory we have presented the classical PV and proposed a new time stretching algorithm in a unified framework. This approach has allowed us to address and improve on the disadvantages of the classical PV, while preserving the mathematical structure provided by Gabor theory. The proposed algorithm is the first attempt to use non-uniform NSGFs for time-stretching, which allows for a low redundancy of the adaptive TF representation and leads to a fast implementation. The proposed algorithm has been compared to other multi-resolution methods, in a reproducible manner, and we have discussed its advantages and its shortcomings. As a future improvement it could be interesting to connect the techniques presented in this article with the ideas proposed by R{\"o}bel in \cite{Roebel2003}, possibly allowing for an algorithm that uses non-uniform NSGFs without the need for fixing the stretch factor during attack transients.

\section*{Acknowledgment}
We would like to thank the three anonymous reviewers for their suggestions, which clearly improved the over-all presentation of this manuscript. Also a special thanks to Olivier Derrien and Marco Liuni for providing us with the sound files used for comparison.

\ifCLASSOPTIONcaptionsoff
  \newpage
\fi

\bibliographystyle{IEEEtran}

\begin{thebibliography}{10}
\providecommand{\url}[1]{#1}
\csname url@samestyle\endcsname
\providecommand{\newblock}{\relax}
\providecommand{\bibinfo}[2]{#2}
\providecommand{\BIBentrySTDinterwordspacing}{\spaceskip=0pt\relax}
\providecommand{\BIBentryALTinterwordstretchfactor}{4}
\providecommand{\BIBentryALTinterwordspacing}{\spaceskip=\fontdimen2\font plus
\BIBentryALTinterwordstretchfactor\fontdimen3\font minus
  \fontdimen4\font\relax}
\providecommand{\BIBforeignlanguage}[2]{{%
\expandafter\ifx\csname l@#1\endcsname\relax
\typeout{** WARNING: IEEEtran.bst: No hyphenation pattern has been}%
\typeout{** loaded for the language `#1'. Using the pattern for}%
\typeout{** the default language instead.}%
\else
\language=\csname l@#1\endcsname
\fi
#2}}
\providecommand{\BIBdecl}{\relax}
\BIBdecl

\bibitem{Ishizaki2009}
H.~Ishizaki, K.~Hoashi, and Y.~Takishima, ``Full-automatic dj mixing system
  with optimal tempo adjustment based on measurement function of user
  discomfort.'' in \emph{ISMIR}.\hskip 1em plus 0.5em minus 0.4em\relax
  International Society for Music Information Retrieval, 2009, pp. 135--140.

\bibitem{Risset20022}
J.-C. Risset, ``Examples of the musical use of digital audio effects,''
  \emph{Journal of New Music Research}, vol.~31, no.~2, pp. 93--97, 2002.

\bibitem{Flanagan1966}
J.~L. Flanagan and R.~M. Golden, ``Phase vocoder,'' \emph{The Bell System
  Technical Journal}, vol.~45, no.~9, pp. 1493--1509, Nov 1966.

\bibitem{Portnoff1976}
M.~Portnoff, ``{Implementation of the digital phase vocoder using the fast
  Fourier transform},'' \emph{IEEE Transactions on Acoustics, Speech, and
  Signal Processing}, vol.~24, no.~3, pp. 243--248, Jun 1976.

\bibitem{Griffen1984}
D.~Griffin and J.~Lim, ``{Signal estimation from modified short-time Fourier
  transform},'' \emph{IEEE Transactions on Acoustics, Speech, and Signal
  Processing}, vol.~32, no.~2, pp. 236--243, Apr 1984.

\bibitem{Laroche1995}
E.~Moulines and J.~Laroche, ``Non-parametric techniques for pitch-scale and
  time-scale modification of speech,'' \emph{Speech Commun.}, vol.~16, no.~2,
  pp. 175--205, feb 1995.

\bibitem{Laroche1999}
J.~Laroche and M.~Dolson, ``Improved phase vocoder time-scale modification of
  audio,'' \emph{IEEE Transactions on Speech and Audio Processing}, vol.~7,
  no.~3, pp. 323--332, May 1999.

\bibitem{Roebel2003}
\BIBentryALTinterwordspacing
A.~R{\"o}bel, ``{A new approach to transient processing in the phase
  vocoder},'' in \emph{{6th International Conference on Digital Audio Effects
  (DAFx)}}, London, United Kingdom, Sep. 2003, pp. 344--349. [Online].
  Available: \url{https://hal.archives-ouvertes.fr/hal-01161124}
\BIBentrySTDinterwordspacing

\bibitem{Dorran2004}
D.~Dorran and R.~Lawlor, ``An efficient phasiness reduction technique for
  moderate audio time-scale modification,'' in \emph{in Proc. of the 7 th
  International Conference on Digital Audio Effects (DAFx-04)}, 2004.

\bibitem{Bonada2000}
J.~Bonada, ``Automatic technique in frequency domain for near-lossless
  time-scale modification of audio,'' in \emph{Proceedings of International
  Computer Music Conference}, 2000, pp. 396--399.

\bibitem{Derrien2007}
\BIBentryALTinterwordspacing
O.~Derrien, ``{Time-scaling of audio signals with multi-scale Gabor
  analysis},'' in \emph{{DAFx'07}}, Bordeaux, France, Sep. 2007, pp. CD--ROM (6
  pages). [Online]. Available:
  \url{https://hal.archives-ouvertes.fr/hal-00467531/file/Derrien_DAFx07.pdf}
\BIBentrySTDinterwordspacing

\bibitem{Liuni2013}
M.~Liuni, A.~R{\"o}bel, E.~Matusiak, M.~Romito, and X.~Rodet, ``Automatic
  adaptation of the time-frequency resolution for sound analysis and
  re-synthesis,'' \emph{IEEE Transactions on Audio, Speech, and Language
  Processing}, vol.~21, no.~5, pp. 959--970, 2013.

\bibitem{Christensen2016}
\BIBentryALTinterwordspacing
O.~Christensen, \emph{An introduction to frames and {R}iesz bases}, 2nd~ed.,
  ser. Applied and Numerical Harmonic Analysis.\hskip 1em plus 0.5em minus
  0.4em\relax Birkh\"auser/Springer, [Cham], 2016. [Online]. Available:
  \url{http://dx.doi.org/10.1007/978-3-319-25613-9}
\BIBentrySTDinterwordspacing

\bibitem{Grochenig2001}
\BIBentryALTinterwordspacing
K.~Gr\"ochenig, \emph{Foundations of time-frequency analysis}, ser. Applied and
  Numerical Harmonic Analysis.\hskip 1em plus 0.5em minus 0.4em\relax
  Birkh\"auser Boston, Inc., Boston, MA, 2001. [Online]. Available:
  \url{http://dx.doi.org/10.1007/978-1-4612-0003-1}
\BIBentrySTDinterwordspacing

\bibitem{NuHAGnsgf2011}
\BIBentryALTinterwordspacing
P.~Balazs, M.~D{\"o}rfler, F.~Jaillet, N.~Holighaus, and G.~Velasco, ``Theory,
  implementation and applications of nonstationary {G}abor frames,'' \emph{J.
  Comput. Appl. Math.}, vol. 236, no.~6, pp. 1481--1496, 2011. [Online].
  Available:
  \url{http://www.sciencedirect.com/science/article/pii/S0377042711004900}
\BIBentrySTDinterwordspacing

\bibitem{Dorfler2014}
M.~D{\"o}rfler and E.~Matusiak, ``{Nonstationary Gabor frames - existence and
  construction},'' \emph{International Journal of Wavelets, Multiresolution and
  Information Processing}, vol.~12, no.~03, p. 1450032, 2014.

\bibitem{Roucos1985}
S.~Roucos and A.~Wilgus, ``High quality time-scale modification for speech,''
  in \emph{ICASSP '85. IEEE International Conference on Acoustics, Speech, and
  Signal Processing}, vol.~10, Apr 1985, pp. 493--496.

\bibitem{Charpentier1986}
F.~Charpentier and M.~Stella, ``Diphone synthesis using an overlap-add
  technique for speech waveforms concatenation,'' in \emph{ICASSP '86. IEEE
  International Conference on Acoustics, Speech, and Signal Processing},
  vol.~11, Apr 1986, pp. 2015--2018.

\bibitem{Puckette1995}
M.~Puckette, ``Phase-locked vocoder,'' in \emph{Proceedings of 1995 Workshop on
  Applications of Signal Processing to Audio and Accoustics}, Oct 1995, pp.
  222--225.

\bibitem{Dixon06}
S.~Dixon, ``Onset detection revisited,'' in \emph{Proc. of the Int. Conf. on
  Digital Audio Effects (DAFx-06)}, Montreal, Quebec, Canada, sep 2006, pp.
  133--137.

\bibitem{Beauregard2015}
\BIBentryALTinterwordspacing
G.~T. Beauregard, M.~Harish, and L.~Wyse, ``Single pass spectrogram
  inversion,'' in \emph{2015 IEEE International Conference on Digital Signal
  Processing (DSP)}, 2015, pp. 427--431. [Online]. Available:
  \url{http://ieeexplore.ieee.org/stamp/stamp.jsp?tp=&arnumber=7251907&isnumber=7251315}
\BIBentrySTDinterwordspacing

\bibitem{Strohmer1998}
\BIBentryALTinterwordspacing
T.~Strohmer, ``Numerical algorithms for discrete {G}abor expansions,'' in
  \emph{Gabor analysis and algorithms}, ser. Appl. Numer. Harmon. Anal.\hskip
  1em plus 0.5em minus 0.4em\relax Birkh\"auser Boston, Boston, MA, 1998, pp.
  267--294. [Online]. Available:
  \url{http://dx.doi.org/10.1007/978-1-4612-2016-9_9}
\BIBentrySTDinterwordspacing

\bibitem{Sondergaard2007}
P.~L. S{\o}ndergaard, ``{Finite discrete Gabor analysis},'' Ph.D. dissertation,
  Technical University of Denmark, 2007.

\bibitem{Daubechies1986}
I.~{Daubechies}, A.~{Grossmann}, and Y.~{Meyer},
  ``\BIBforeignlanguage{English}{{Painless nonorthogonal expansions.}}''
  \emph{\BIBforeignlanguage{English}{{J. Math. Phys.}}}, vol.~27, pp.
  1271--1283, 1986.

\bibitem{Robel2013}
M.~Liuni and A.~R{\"o}bel, ``Phase vocoder and beyond,''
  \emph{Musica/Tecnologia}, vol.~7, pp. 73--89, 2013.

\bibitem{Laroche2002}
\BIBentryALTinterwordspacing
J.~Laroche, \emph{Time and Pitch Scale Modification of Audio Signals}.\hskip
  1em plus 0.5em minus 0.4em\relax Boston, MA: Springer US, 2002, pp. 279--309.
  [Online]. Available: \url{http://dx.doi.org/10.1007/0-306-47042-X_7}
\BIBentrySTDinterwordspacing

\bibitem{McAulay1986}
R.~McAulay and T.~Quatieri, ``Speech analysis/synthesis based on a sinusoidal
  representation,'' \emph{IEEE Transactions on Acoustics, Speech, and Signal
  Processing}, vol.~34, no.~4, pp. 744--754, Aug 1986.

\bibitem{ltfatnote030}
\BIBentryALTinterwordspacing
Z.~Pr\r{u}\v{s}a, P.~L. S{\o}ndergaard, N.~Holighaus, C.~Wiesmeyr, and
  P.~Balazs, ``The large time-frequency analysis toolbox 2.0,'' in \emph{Sound,
  Music, and Motion}, ser. Lecture Notes in Computer Science.\hskip 1em plus
  0.5em minus 0.4em\relax Springer International Publishing, 2014, pp.
  419--442. [Online]. Available:
  \url{{http://dx.doi.org/10.1007/978-3-319-12976-1_25}}
\BIBentrySTDinterwordspacing

\bibitem{Ellis2002}
\BIBentryALTinterwordspacing
D.~P.~W. Ellis, ``A phase vocoder in {M}atlab,'' 2002, web resource. [Online].
  Available: \url{http://www.ee.columbia.edu/~dpwe/resources/matlab/pvoc/}
\BIBentrySTDinterwordspacing
\end{thebibliography}

\end{document}